\pdfoutput = 1
\documentclass[journal=nalefd, manuscript=letter]{achemso}
\usepackage{graphicx}
\usepackage{graphics}
\usepackage{amsmath}
\usepackage{amssymb}
\usepackage{amsfonts}
\usepackage{dcolumn}
\usepackage{dsfont}
\usepackage{latexsym}
\usepackage{rotating}
\usepackage{array}
\usepackage{makecell}
\usepackage{color}
\usepackage{latexsym}
\usepackage{bbm}
\usepackage{physics}
\usepackage{subfigure}
\usepackage{float}
\usepackage{epsfig}
\usepackage{epsf}
\usepackage{psfrag}
\usepackage{bm}
\usepackage{amsthm}
\usepackage{eucal}
\usepackage{mathrsfs}
\usepackage{tabularx}
\usepackage{url}
\usepackage{braket}
\usepackage{epstopdf}
\usepackage{color} 

\usepackage{hyperref}
\hypersetup{
colorlinks=true,
        final=true,
        linkcolor=black,
        citecolor=black,
        filecolor=black,
        urlcolor=black,
}

\title{Theory of moir\'e magnetism in twisted bilayer $\alpha$-RuCl$_3$}

\author{Muhammad Akram}
\affiliation{Department of Physics, Arizona State University, Tempe, AZ 85287, USA}
\author{Jesse Kapeghian}
\affiliation{Department of Physics, Arizona State University, Tempe, AZ 85287, USA}
\author{Jyotirish Das}
\affiliation{Department of Physics, Arizona State University, Tempe, AZ 85287, USA}
\author{Roser Valent\'\i}
\affiliation{Institut f\"ur Theoretische Physik, Goethe-Universit\"at Frankfurt, 60438 Frankfurt am Main, Germany}
\author{Antia S. Botana}
\affiliation{Department of Physics, Arizona State University, Tempe, AZ 85287, USA}
\email{antia.botana@asu.edu}
\author{Onur Erten}
\affiliation{Department of Physics, Arizona State University, Tempe, AZ 85287, USA}
\email{onur.erten@asu.edu}

\begin{document}

KEYWORDS: 2D vdW magnets, $\alpha$-RuCl$_3$, moir\'e patterns.

\begin{abstract}
Twisted heterostructures of van der Waals materials have received much attention for their many remarkable properties. Here, we present a comprehensive theory of the long-range ordered magnetic phases of twisted bilayer $\alpha$-RuCl$_3$ via a combination of first-principles calculations and atomistic simulations. While a monolayer exhibits zigzag antiferromagnetic order with three possible ordering wave vectors, a rich phase diagram is obtained for moir\'e superlattices as a function of interlayer exchange and twist angle. For large twist angles, each layer spontaneously picks a single zigzag ordering wave vector, whereas, for small twist angles, the ground state involves a combination of all three wave vectors in a complex hexagonal domain structure. This multi-domain order minimizes the interlayer energy while enduring the energy cost due to the domain wall formation. Our results indicate that magnetic frustration due to stacking-dependent interlayer exchange in moir\'e superlattices can be used to tune the magnetic ground state and enhance quantum fluctuations in $\alpha$-RuCl$_3$. 

\end{abstract}


$\alpha$-RuCl$_3$ is a van der Waals (vdW) magnet that has attracted a
lot of attention in recent years as a promising candidate for  a Kitaev quantum spin liquid~\cite{kitaev2006anyons}. However,  its zigzag antiferromagnetic (AFM) order below T$_N$ $\sim$ 7 K \cite{Banerjee_Science2017, Banerjee_NatMat2016, Cao_PRB2016,plumb_PRB2014,johnson_PRB2015} indicates deviations from the Kitaev model.  The
specific nature of these deviations has been theoretically and experimentally intensively scrutinized, with many works pointing to additional large anisotropic couplings beyond the Kitaev interaction and long-range exchange likely stabilizing magnetic order~\cite{Rau_PRL2014,Winter_PRB2016,Winter_NatComm2017,winter2017_review,Maksimov_PRR2020,rousochatzakis2023_review}. 
Under application of an in-plane magnetic field the zigzag order is suppressed leading to a  field-induced `disordered' phase, whose  nature  is presently still under intensive debate, specially due to initial reports of the observation of half-quantized thermal Hall conductivity~\cite{Yokoi_Science2021}. This behavior  would correspond to a chiral Kitaev spin liquid state~\cite{kitaev2006anyons}. However, these results have
not been corroborated so far~\cite{lefranccois2022,Bruin_NatPhys2022,czajka_NatMat2023}.
Alternatively, there have been promising routes to investigate the emergence of novel phases beyond the zigzag AFM order  by considering such approaches as  chemical doping~\cite{bastien_PRB2019}, strain fields~\cite{bastien2018,bachus2020,kocsis2022,kaib2021}  and graphene substrates~\cite{mashhadi2019,zhou2019,biswas2019,Rizzo_NanoLett2020,gerber2020,leeb_PRL2021,balgley2022,shi2023magnetic}. 
In the present work, we explore yet a new route to modify the magnetism in $\alpha$-RuCl$_3$, by exploring the properties of twisted bilayers of $\alpha$-RuCl$_3$. 

Moir\'e superlattices of van der Waals materials have surfaced as new tunable quantum platforms for the realization of emergent phases on every front including graphene\cite{Cao2018}, semiconductors\cite{Devakul_NatComm2021}, and superconductors\cite{Zhao_arxiv2021}. While moir\'e engineering of electronic phases has been studied extensively, research in moir\'e superlattices comprised of magnetic materials is at its early stages. A range of novel non-coplanar phases have been predicted in moir\'e vdW magnets\cite{Tong_ACS2018, Hejazi_PNAS2020, Hejazi_PRB2021, Akram_PRB2021, Akram_NanoLett2021, Tong_PRR2021, Xiao_PRR2021, Ghader_CommPhys2022, Zheng_AFM2023, Kim_NanoLett2023, Fumega_2DMAtt2023} and some of these phases have been observed experimentally\cite{Xu_NatNano2021, Song_Science2021, xie2022}. Given that monolayers of $\alpha$-RuCl$_3$ can be isolated by exfoliation methods\cite{Lee_npjQM2021, Yang_NatMat2023} and heterostructures with other vdW materials such as graphene can be constructed\cite{Rizzo_NanoLett2020},  moir\'e engineering in $\alpha$-RuCl$_3$ can be explored as a means to tune its magnetism.

Here, we study the long-range ordered magnetic phases of twisted bilayer $\alpha$-RuCl$_3$ by a combination of first-principles calculations and atomistic simulations. Our main results are as follows: (i) we obtain the stacking dependent interlayer exchange within the moir\'e unit cell and show that the two layers are coupled antiferromagnetically for different stacking orders. (ii) Among the three inequivalent ordering wave vectors ($q_i$, $i = 1,~2,~3$) for the zigzag AFM order (see Fig.~\ref{Fig2}), the two layers spontaneously pick different $q$'s on each layer for large twist angles. This single-$q$ phase ($1q-1q$) has no domain wall but the interlayer exchange energy averages to almost zero. (iii) For small angles or large moir\'e periodicity, the ground state incorporates all of the three $q$'s in a complex domain structure ($3q-3q$). These domains resemble a hexagonal shape and minimize the interlayer exchange. However, these domains are separated by domain walls that cost energy. (iv) Comparing the interlayer exchange and domain wall energy, we obtain an analytical formula for the transition between these two phases which agrees well with the atomistic simulations. (v) In the vicinity of the ($1q-1q$) and ($3q-3q$) phase boundary, we obtain an intermediate phase with two $q$'s on each layer ($2q-2q$) that has both domain walls and finite interlayer exchange energy.

We start by introducing the spin Hamiltonian which describes the magnetic properties of twisted bilayer  $\alpha$-RuCl$_3$.  For simplicity we limit ourselves to a minimal set of exchange parameters, which have proven to provide a reasonable description of several experimental observations in $\alpha$-RuCl$_3$~\cite{Winter_NatComm2017}:
\begin{equation}
\label{eqn:spinH}
\mathcal{H} = \mathcal{H}_{intra}^1+\mathcal{H}_{intra}^2+\mathcal{H}_{inter},
\end{equation}
where $\mathcal{H}_{intra}^{1 (2)}$ contains the intralayer exchange terms in layer 1 (2) and $\mathcal{H}_{inter}$ contains the interlayer exchange, 
\begin{eqnarray} 
\label{Eq:Gnrl_Hmlt_1}
\mathcal{H}_{intra}&=&\sum_{\langle ij \rangle_\gamma}\big[ J_1 \mathbf{S}_{i}\cdot \mathbf{S}_{j}+ K S_i^\gamma S_j^\gamma + \Gamma (S_i^\alpha S_j^\beta+S_j^\alpha S_i^\beta) \big] + J_3 \sum_{\langle \langle \langle ij \rangle \rangle \rangle} \mathbf{S}_{i}\cdot \mathbf{S}_{j},\\
\label{eq:3}
\mathcal{H}_{inter}&=& \sum_{\langle ij \rangle} J_\perp({ r}_{ij}) {\bf S}_i^{1}\cdot {\bf S}_j^{2}.
\end{eqnarray}
where $\gamma = x~,y,~z$ type bonds on the honeycomb lattice. $J_{1(3)}$ is the (third) nearest-neighbor Heisenberg exchange, $K$ and $\Gamma$ are the Kitaev and the symmetric anisotropic exchange interaction terms\cite{Rau_PRL2014, Winter_PRB2016}. $J_\perp({r}_{ij})$ is the interlayer exchange coupling and ${ r}_{ij}$ represents the interlayer displacement. 

\begin{figure}[t]
\center
\includegraphics[width=\textwidth]{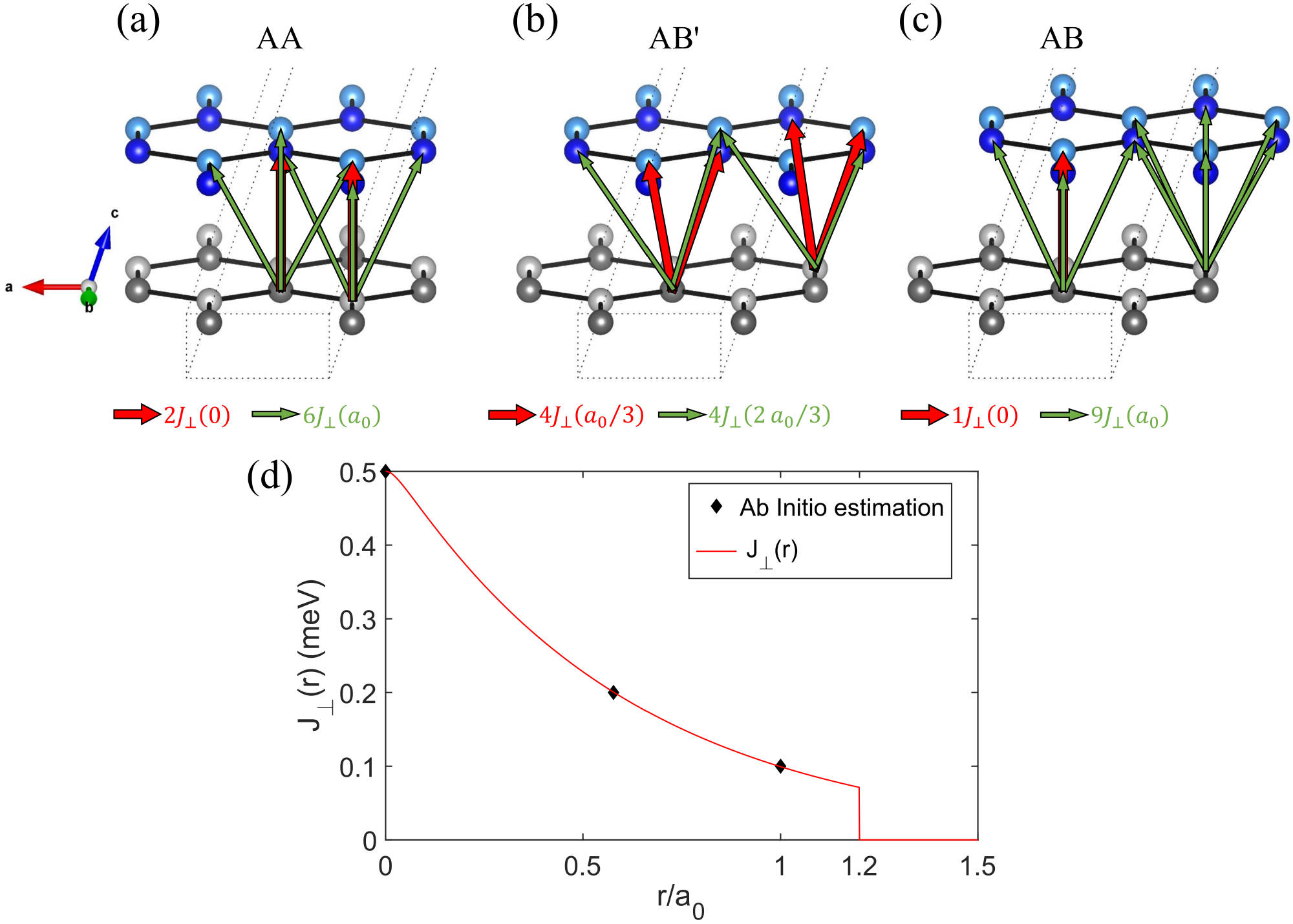} 
\caption{(a-c) Schematic representation of the three different stackings used in our first-principles DFT calculations displaying the interlayer neighbors exchange couplings considered (for half a unit cell) as shown in Eqs. 4-6. (d) Continuous interlayer exchange coupling function $J_{\perp}(r)=J_\perp(0)e^{\frac{-B\sqrt{C^2+r^2}}{C}}/e^{-B}$ as a function of displacement $r$. Here, we use $J_{\perp}(0)=0.5$ meV obtained from  our ab initio DFT estimation.}
\label{Fig1}
\end{figure}

\begin{figure}[t]
\center
\includegraphics[width=\textwidth]{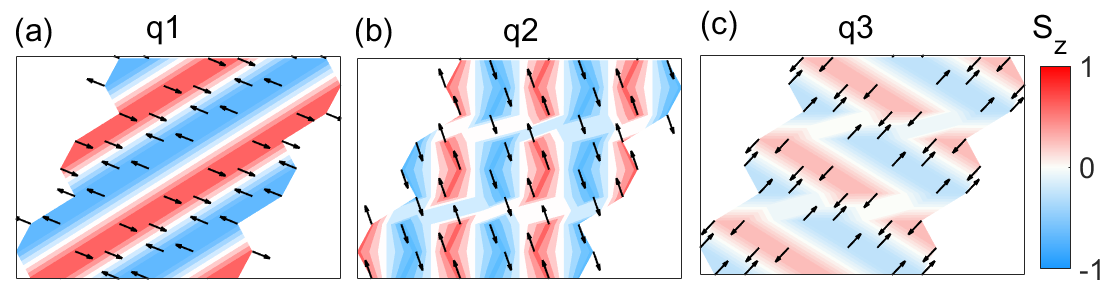} 
\caption{Spin textures for the zigzag AFM order for a $7\times7$ system. There are three inequivalent ordering wave vectors (a) $q_1$, (b) $q_2$ and (c) $q_3$ that are related with $C_3$ symmetry. 
}
\label{Fig2}
\end{figure}

For the intralayer exchange parameters, we used $K= -5$ meV, $\Gamma= 2.5$ meV, $J_{1}$= -0.5 meV, and $J_3$ = 0.5 meV as obtained in Ref.~\citenum{Winter_NatComm2017}.
In order to determine the interlayer exchange, we perform first-principles density functional theory (DFT) calculations for  three different bilayer stackings (AA, AB', and AB) shown in Fig. \ref{Fig1} (details on the \textit{ab initio} DFT calculations are provided in the Supporting Information). For each stacking, we extract the energy difference ($\Delta$$E$) between two spin configurations: ferromagnetic (FM), with FM planes coupled FM out-of-plane, and antiferromagnetic (AFM), corresponding to FM planes coupled AFM out-of-plane. Even though the magnetic ground state in RuCl$_3$ is  zig-zag-like, as mentioned above, we use the energy difference between these simpler spin configurations to obtain an estimate of the effective interlayer exchange. Importantly, for all stackings we obtain a lower energy for an effective AFM interlayer coupling. 

For the three stacking orders used, we construct an interlayer spin hamiltonian in which the exchange coupling depends on the displacement, $J_\perp(r_{ij})$ and we equate the energy difference from the {\it ab initio} DFT alculations ($\Delta$$E$) to the corresponding effective interlayer spin Hamiltonian. For instance, the unit cell for the AA stacking used in the first-principles calculations has 4 interlayer bonds with $r_{ij}=0$ and 12 bonds with $r_{ij}=a/\sqrt{3}$ (see Fig. \ref{Fig1}(a-c) for a schematic picture of the interlayer bonds). In this manner, we obtain three equations for the three different stacking patterns used.  
\begin{eqnarray}
    4J_\perp(0)+12J_\perp(a_0)&=&\Delta E_{AA}/2=3.2~ {\rm meV} \label{eq:EAA} \\
    8 J_\perp(a_0/3)+8J_\perp(2a_0/3)&=&\Delta E_{AB'}/2=1.6~ {\rm meV} \label{eq:EABp}\\ 
    2 J_\perp(0)+18 J_\perp(a_0)&=&\Delta E_{AB}/2=2.8~ {\rm meV} \label{eq:EAB}
\end{eqnarray}
where $a_0=a/\sqrt{3}$ is the bond length on the honeycomb lattice.  We implement a cut-off for the range of the exchange interaction, $J(r=2a/3)\ge 0$ and solve the Eqs. \ref{eq:EAA}, \ref{eq:EABp} and \ref{eq:EAB}, obtaining $J(0)=0.5$ meV, $J(a_0/\sqrt{3})=0.2$ meV and $J(a_0)=0.1$ meV. Similar values of the interlayer exchange parameters have been reported in previous work \cite{Janssen_RPB2020, Balz_PRB2021}. To determine a continuous $J(r_{ij})$, we fit these values to an exponential decaying function, $J_\perp(r)=J_\perp(0)e^{\frac{-B\sqrt{C^2+r^2}}{C}}/e^{-B}$ (as shown in  Fig.~\ref{Fig1}(d)) with $B=0.04898$ and $C=0.02943$ being the fitting parameters.

\begin{figure}
\center
\includegraphics[width=\textwidth]{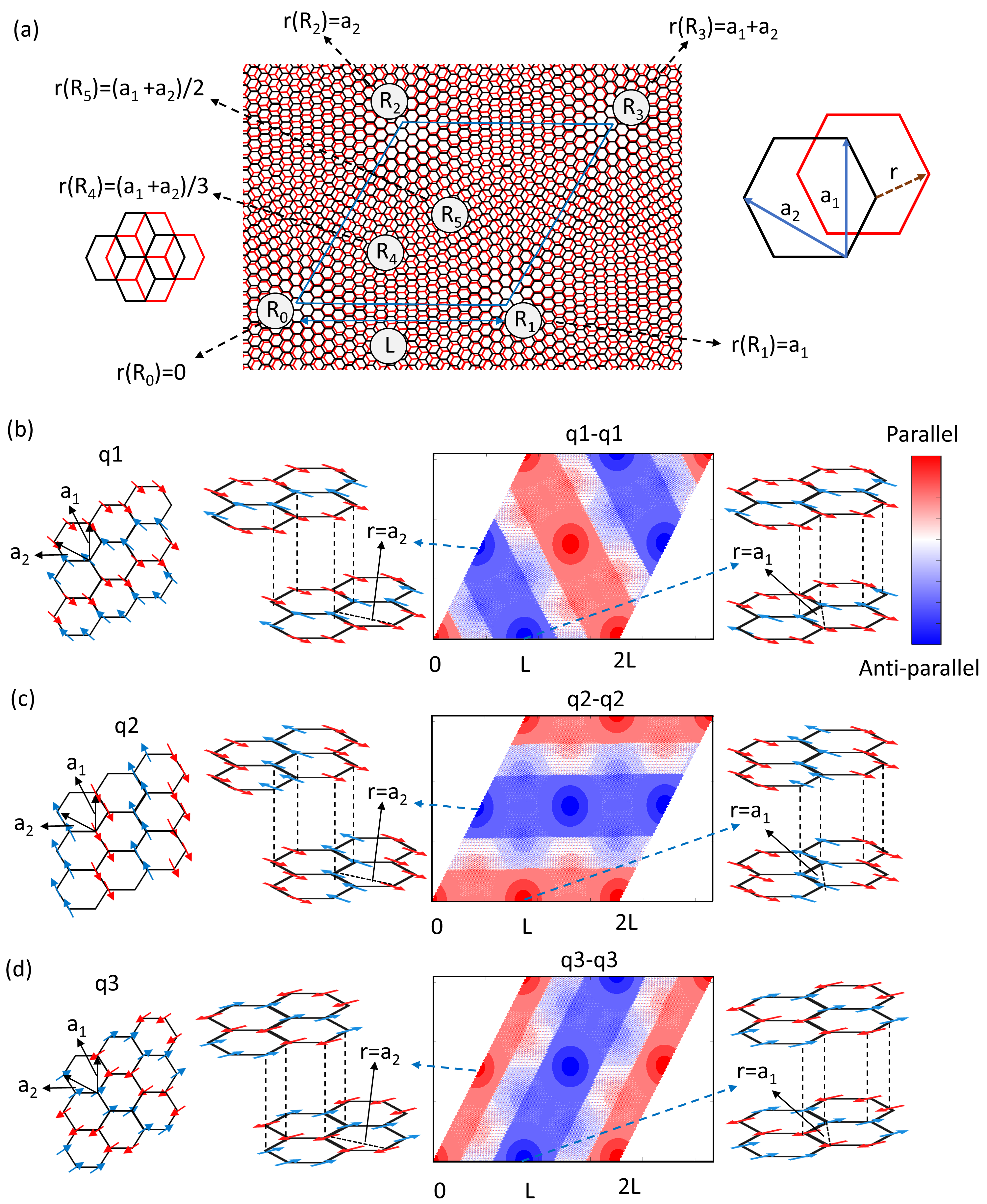} 
\caption{Moir\'e pattern and projection of the nearest neighbor spins of layer-2 on the spins of layer-1 in twisted bilayer RuCl$_3$. (a) Moir\'e pattern of a twisted bilayer system. For small angle the local stackings of the moir\'e unit cell can be considered as the untwisted bilayer system with one layer shifted by displacement $r$. (b) $q_1$ zigzag order and the projection of first nearest neighbour spins of layer-2 on the spins of layer-1 times an exponential decaying function of in-plane displacement $r$ with $q_1$ zigzag order in both layers of a 2$\times$2 moir\'e unit cell, ($L\sim a/\theta$). The red and blue colors represent parallel and anti-parallel spins respectively. (c) and (d) represent the same analysis as (b) for $q_2$ and $q_3$, respectively.}
\label{Fig:NN}
\end{figure}

For a bilayer `untwisted' $\alpha$-RuCl$_3$ with AA stacking order, the antiferromagnetic interlayer exchange is not frustrated and the energy is minimized when both layers have the same single-$q$ zigzag order. Due to the three-fold rotational symmetry, there are three possible single-$q$ patterns as shown in Fig.~\ref{Fig2}. On the contrary, for moir\'e superlattices, the real space stacking within the moir\'e unit cell changes as a function of displacement $\bf{R}$. For small twist angles, the local stacking can by described by a translational shift, ${\bf r}(\bf{R}) \simeq  \theta \hat{z} \times {\bf R}$ as shown in Fig.~\ref{Fig:NN} (a). Due to this varying shift, the nearest neighbour spins in two layers cannot be anti-parallel throughout the moir\'e unit cell when both layers have the same single-$q$ order. To show this effect, we present  in Fig.~\ref{Fig:NN} (b),(c) and (d) the projections of the nearest neighbour spins of layer-2 on the spins of layer-1 times the exponential decaying function $e^{\frac{-B\sqrt{C^2+r^2}}{C}}/e^{-B}$ used above for $J_\perp$ for $q_1$, $q_2$, and $q_3$ respectively. For $(1q-1q)$ zigzag order with $q_1$ on each layer, the spins are parallel at $R_0$ with a zero shift as shown in Fig.~\ref{Fig:NN} (b). At $R_1$ and $R_2$, the translational shifts are $a_1$ and $a_2$, respectively, and both of these shifts move a parallel spin to the location of an anti-parallel spin. The whole pattern of varying parallel and anti-parallel local stacking orders can be obtained in this way for the $q_2$ and $q_3$ orders as well. Based on the local translational shifts, the ($1q-1q$) zigzag order in a $2\times2$ moir\'e unit cell can be divided into three equal parts as shown in Fig.~\ref{Fig:NN} (b), (c), and (d). In one-third area, the spins are aligned parallel to each other and this area is represented by red color, whereas in the second one-third area, they are arranged anti-parallel (blue color). In the last one-third area, the spins are both parallel and anti-parallel (white color). Therefore, the total interlayer exchange energy averages out to zero. Even though this analysis is carried out when both layers have the same wave vector, it holds true for different $q_i$'s on each layer as well.

\begin{figure}[t]
\center
\includegraphics[width=1\textwidth]{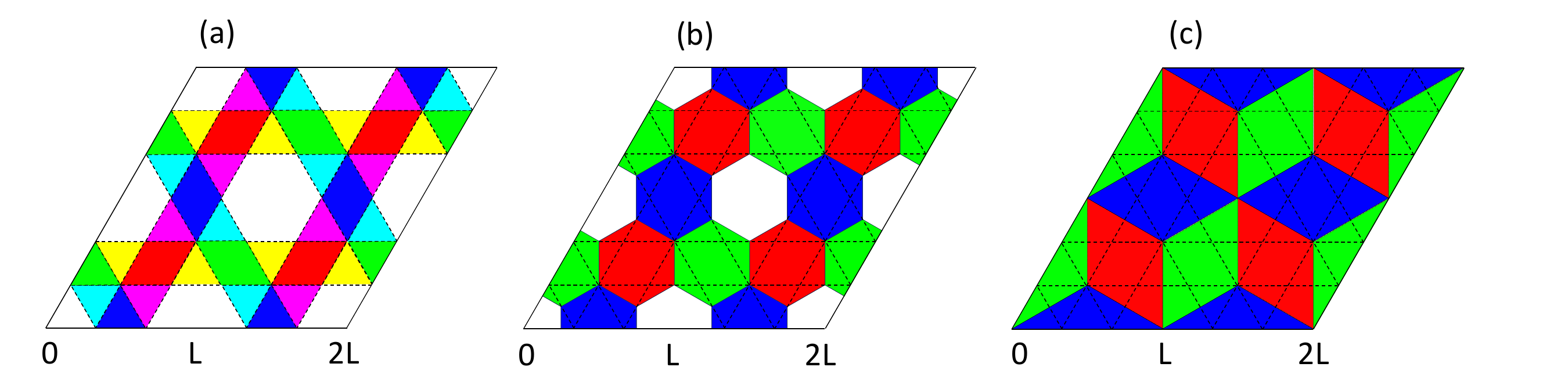} 
\caption{(a) Schematic representation of constraints on the wave vectors that minimize the interlayer exchange energy. Red, blue and green diamonds represent regions where only $q_1$, $q_2$ and $q_3$ can gain interlayer energy respectively. On the other hand, violet, yellow and pink colors represent regions where  $q_1$, $q_2$ and $q_3$ on their own cannot gain energy. White color represent regions where all the three $q$'s can gain interlayer energy. (b,c) Two examples of multi-domain structures that minimize the interlayer exchange energy. Red, blue and green colors represent $q_1$, $q_2$ and $q_3$ zigzag orders respectively. The white hexagons in (b) represent regions where any of the three $q$'s is allowed.}
\label{Fig:3q}
\end{figure}

However, the interlayer energy can be minimized if there are multiple-$q$ zigzag patterns in each layer that are separated by domain walls. In Fig.~\ref{Fig:NN} (b), (c) and (d), the white regions represent the areas where $q_1$, $q_2$ and $q_3$ orders cannot gain interlayer energy, respectively. These regions are combined in a single graph in Fig.~\ref{Fig:3q}(a) that shows a schematic representation of constraints on the wave vectors that minimize the interlayer exchange energy. We implement the following color coding: the intersection of yellow and magenta bars is represented by red diamonds where both $q_2$ and $q_3$ orders cannot gain energy. Therefore, the red diamonds display $q_1$ order. Similarly, blue and green diamonds represent regions where only $q_2$ and $q_3$ can gain energy respectively. The violet, yellow and pink colors represent regions where only $q_1$, $q_2$ and $q_3$ cannot gain energy. In the white region, any of the three wave vectors can gain energy. Applying these rules, we have constructed two examples of multi-domain structures in Fig.~\ref{Fig:3q} (b) and (c). The white hexagons in Fig.~\ref{Fig:3q} (b) can be filled with any of the three $q$ orders. The domain structure in Fig.~\ref{Fig:3q}(c) was predicted in Ref.~\citenum{Hejazi_PNAS2020} using a continuum model instead. Both of these configurations ((b) and (c)) fully minimize the interlayer energy and only differ by the domain wall length.

\begin{figure}[t]
\center
\includegraphics[width=0.9\textwidth]{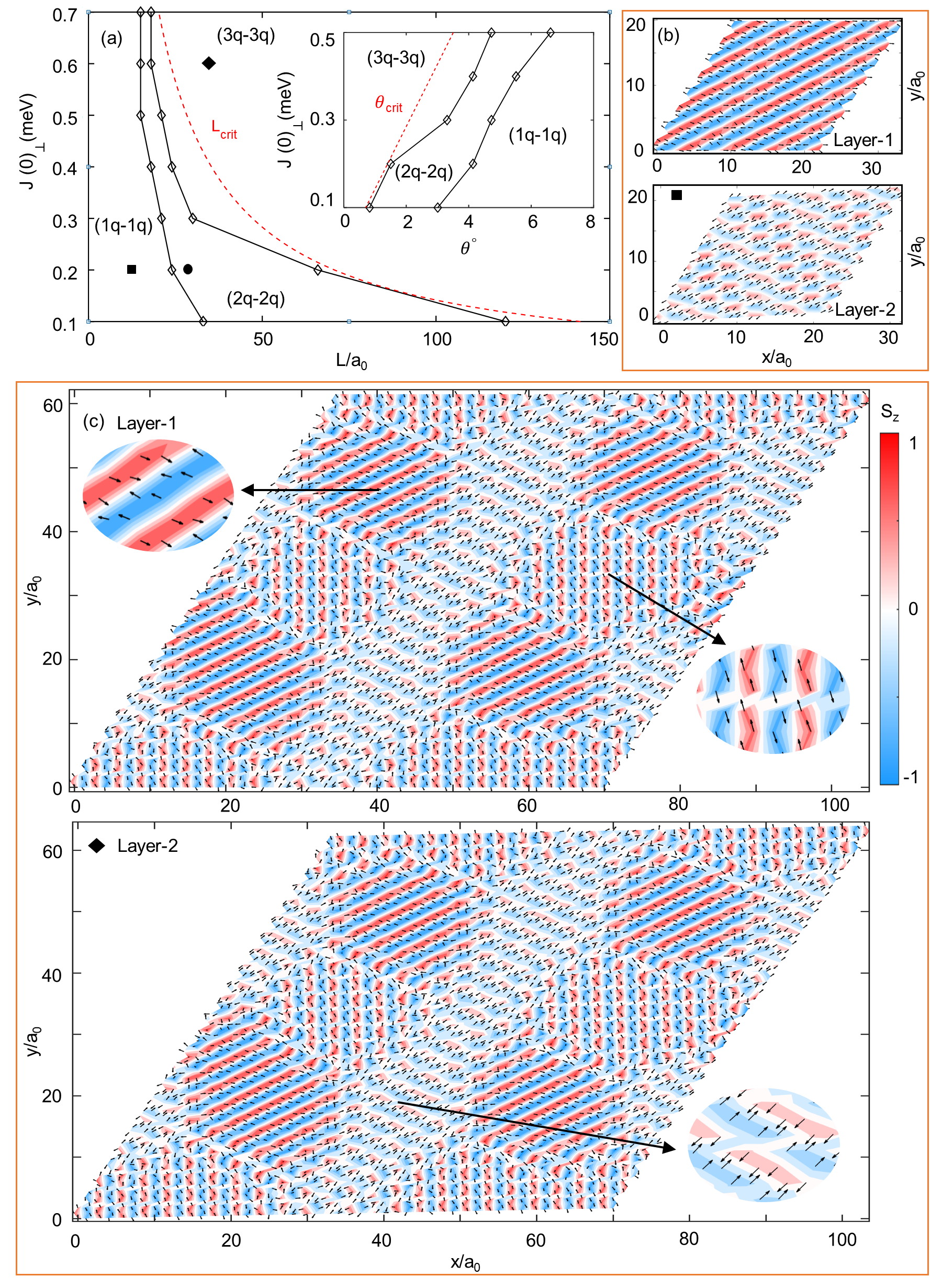} 
\caption{(a) Phase diagram of twisted bilayer $\alpha$-RuCl$_3$ as a function moir\'e period $L$ and $J_{\perp}$. The {\it ab initio} estimate is $J_\perp (0) = 0.5$ meV. The inset shows the same phase diagram as a function twisted angle $\theta$ and interlayer coupling $J_{\perp}$. Magnetization texture of (b) $(1q-1q)$ for $L=12a_0$ and $J_{\perp}(0)=0.2$ meV; and (c) $(3q-3q)$ for $L=36a_0$ and $J_{\perp}(0)=0.6$ meV. Magnetization textures are shown for $2\times2$ moir\'e unit cells.}
\label{Fig:5}
\end{figure}

\begin{figure}[t]
\center
\includegraphics[width=\textwidth]{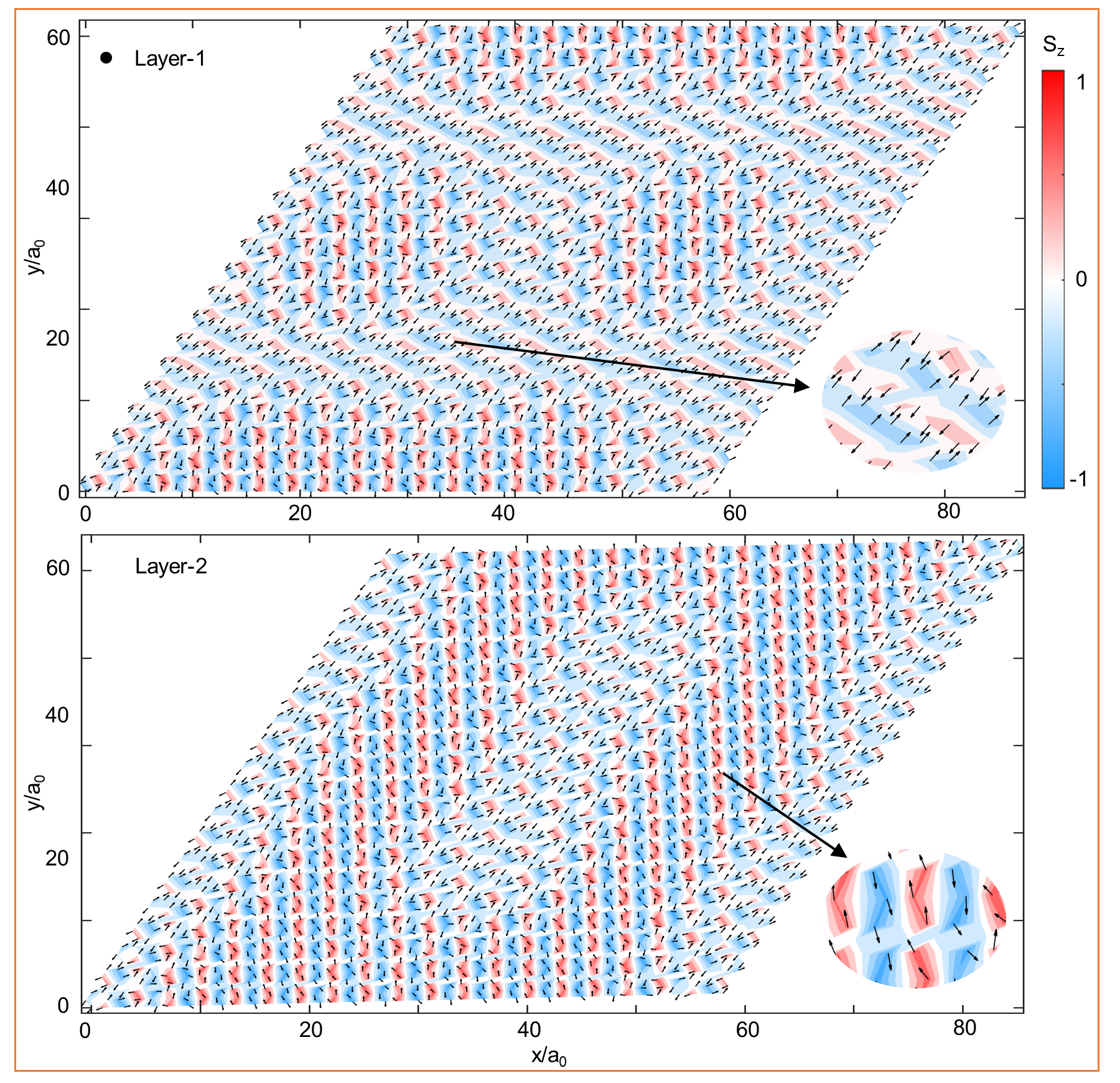} 
\caption{Magnetization texture of $(2q-2q)$ for $L=30a_0$ and $J_{\perp}(0)=0.2$ meV. Magnetization texture is shown for $2\times2$ moir\'e unit cells.}
\label{Fig:6}
\end{figure}

The competition between the domain wall and the interlayer energy leads to interesting phases as a function of the moir\'e period (or twist angle) and interlayer coupling. We obtain the corresponding phase diagram by solving the Landau-Lifshitz-Gilbert (LLG) equation\cite{1353448} for both layers as shown in Fig.~\ref{Fig:5}(a). For the interlayer exchange, we use the form derived from first-principles calculations but we vary the overall amplitude, $J_\perp(0)$. For small moir\'e period (or large twist angle), the ground state is in a ($1q-1q)$ phase as shown in in Fig.~\ref{Fig:5}(b). In this phase, both layers have different single-$q$ zigzag patterns. These zigzag patterns are deformed from perfect order, leading to some gain in interlayer energy. The degree of deformation increases with $J_{\perp}$. For large moir\'e periodicity (L) and large $J_\perp$, we obtain the $(3q-3q)$ phase. In this phase, all three possible zigzag patterns are formed as shown in Fig.~\ref{Fig:5} (c). This phase is consistent with Fig.~\ref{Fig:3q} (b) and the shape of the zigzag patterns is hexagonal. One of the three $q$'s occupies the white hexagons of Fig.~\ref{Fig:3q} (b) and it forms a larger domain of three hexagons connecting the opposite hexagons of the same $q$. This phase maximizes the interlayer energy and the gain in interlayer energy in $2\times2$ moir\'e unit cells is $E_\perp \sim J_\perp \times N \approx J_\perp 8(L/a_0)^2/3$
where $N$ is the number of sites and $a_0$ is the bond length. On the other hand, the cost of domain wall energy formation is approximately $E_{DW} \sim 40L'\times5.6623/\sqrt{3}a_0$ meV where $40L'$ is the length of the domain walls and $L'=L/(4\cos(\pi/6))$ is the edge-length of the hexagons (see Supporting Information for the derivation of $E_{DW}$). A critical moir\'e period for the phase transition from ($1q-1q$) to $(3q-3q)$ can be attained by equating $E_{DW}$ and $E_{\perp}$  leading to $L_{crit}=(14.1562\text{ meV}/J_\perp)a_0$. We find that the second candidate domain structure shown in Fig.~\ref{Fig:3q} (c) has a longer domain wall length, and therefore it is not preferred when compared to the hexagonal domain structure. The dashed line in Fig.~\ref{Fig:5}(a) represents the analytical estimate for the ($1q-1q$) to ($3q-3q$) phase transition. For intermediate moir\'e periods, we obtain an intermediate $(2q-2q)$ phase as shown in Fig.~\ref{Fig:6}. In this phase, there are two kinds of zigzag patterns giving rise to gaining a large fraction of interlayer energy. However, it is not possible to obtain a maximum interlayer energy from double zigzag patterns. On the contrary, such patterns have shorter domain wall length compared to $(3q-3q)$ phase. For small $J_\perp$, the $(2q-2q)$ phase exists for a range of moir\'e periods whereas for large J$_\perp$ this phase exists only for a very short range of small moir\'e periods and the distinction between $(1q-1q)$ and $(2q-2q)$ phases is not evident.

In conclusion, we have shown that the interplay of the stacking-dependent interlayer
exchange and twist angle can play an important role in determining the magnetic phases of $\alpha$-RuCl$_3$. In particular, we demonstrate that the single-domain, $(1q-1q)$ zigzag order can be taken over by multi-$q$ patterns in order to minimize the interlayer exchange energy. These phases appear at small twist angles and can be used as a new route to introduce additional frustration and tune the magnetic phases in $\alpha$-RuCl$_3$. Interesting future directions include estimating the magnon spectrum in multi-$q$ orders and incorporating substrate effects. 

\section{Associated Content}

\noindent {\bf Supporting Information}

\noindent Structural parameters, domain wall energy estimation (pdf).

\section{Author Information}
\noindent {\bf Corresponding Author:}
Onur Erten -- Email: onur.erten@asu.edu

\noindent {\bf Author Contributions:}
Muhammad Akram and Jyotirish Das have performed the atomistic simulations. Jesse Kapeghian has performed the first-principles calculations. Antia Botana and Onur Erten have conceptualized the research and analyzed the results. All authors have contributed to writing the manuscript.

\noindent {\bf Notes:} The authors declare no competing financial interest.

\section{Acknowledgements}
We thank Arun Paramekanti for fruitful discussions. OE and AB acknowledge support from National Science Foundation Award No. DMR 2206987. MA acknowledges support from National Science Foundation Awards No. DMR 2234352. We acknowledge the ASU Research Computing Center for HPC resources. RV acknowledges the DFG (German Research Foundation) for funding through the research unit QUAST FOR 5249 (project ID: 449872909; project P4) and  her research was supported in part by the National Science Foundation under Grants No. NSF PHY-1748958 and PHY-2309135.


\begin{mcitethebibliography}{52}
\providecommand*\natexlab[1]{#1}
\providecommand*\mciteSetBstSublistMode[1]{}
\providecommand*\mciteSetBstMaxWidthForm[2]{}
\providecommand*\mciteBstWouldAddEndPuncttrue
  {\def\EndOfBibitem{\unskip.}}
\providecommand*\mciteBstWouldAddEndPunctfalse
  {\let\EndOfBibitem\relax}
\providecommand*\mciteSetBstMidEndSepPunct[3]{}
\providecommand*\mciteSetBstSublistLabelBeginEnd[3]{}
\providecommand*\EndOfBibitem{}
\mciteSetBstSublistMode{f}
\mciteSetBstMaxWidthForm{subitem}{(\alph{mcitesubitemcount})}
\mciteSetBstSublistLabelBeginEnd
  {\mcitemaxwidthsubitemform\space}
  {\relax}
  {\relax}

\bibitem[Kitaev(2006)]{kitaev2006anyons}
Kitaev,~A. Anyons in an exactly solved model and beyond. \emph{Ann. Phys.}
  \textbf{2006}, \emph{321}, 2--111\relax
\mciteBstWouldAddEndPuncttrue
\mciteSetBstMidEndSepPunct{\mcitedefaultmidpunct}
{\mcitedefaultendpunct}{\mcitedefaultseppunct}\relax
\EndOfBibitem
\bibitem[Banerjee \latin{et~al.}(2017)Banerjee, Yan, Knolle, Bridges, Stone,
  Lumsden, Mandrus, Tennant, Moessner, and Nagler]{Banerjee_Science2017}
Banerjee,~A.; Yan,~J.; Knolle,~J.; Bridges,~C.~A.; Stone,~M.~B.;
  Lumsden,~M.~D.; Mandrus,~D.~G.; Tennant,~D.~A.; Moessner,~R.; Nagler,~S.~E.
  Neutron scattering in the proximate quantum spin liquid
  $\ensuremath{\alpha}\ensuremath{-}{\mathrm{RuCl}}_{3}$. \emph{Science}
  \textbf{2017}, \emph{356}, 1055--1059\relax
\mciteBstWouldAddEndPuncttrue
\mciteSetBstMidEndSepPunct{\mcitedefaultmidpunct}
{\mcitedefaultendpunct}{\mcitedefaultseppunct}\relax
\EndOfBibitem
\bibitem[Banerjee \latin{et~al.}(2016)Banerjee, Bridges, Yan, Aczel, Li, Stone,
  Granroth, Lumsden, Yiu, Knolle, Bhattacharjee, Kovrizhin, Moessner, Tennant,
  Mandrus, and Nagler]{Banerjee_NatMat2016}
Banerjee,~A. \latin{et~al.}  Proximate Kitaev quantum spin liquid behaviour in
  a honeycomb magnet. \emph{Nature Materials} \textbf{2016}, \emph{15},
  733--740\relax
\mciteBstWouldAddEndPuncttrue
\mciteSetBstMidEndSepPunct{\mcitedefaultmidpunct}
{\mcitedefaultendpunct}{\mcitedefaultseppunct}\relax
\EndOfBibitem
\bibitem[Cao \latin{et~al.}(2016)Cao, Banerjee, Yan, Bridges, Lumsden, Mandrus,
  Tennant, Chakoumakos, and Nagler]{Cao_PRB2016}
Cao,~H.~B.; Banerjee,~A.; Yan,~J.-Q.; Bridges,~C.~A.; Lumsden,~M.~D.;
  Mandrus,~D.~G.; Tennant,~D.~A.; Chakoumakos,~B.~C.; Nagler,~S.~E.
  Low-temperature crystal and magnetic structure of
  $\ensuremath{\alpha}\ensuremath{-}{\mathrm{RuCl}}_{3}$. \emph{Phys. Rev. B}
  \textbf{2016}, \emph{93}, 134423\relax
\mciteBstWouldAddEndPuncttrue
\mciteSetBstMidEndSepPunct{\mcitedefaultmidpunct}
{\mcitedefaultendpunct}{\mcitedefaultseppunct}\relax
\EndOfBibitem
\bibitem[Plumb \latin{et~al.}(2014)Plumb, Clancy, Sandilands, Shankar, Hu,
  Burch, Kee, and Kim]{plumb_PRB2014}
Plumb,~K.~W.; Clancy,~J.~P.; Sandilands,~L.~J.; Shankar,~V.~V.; Hu,~Y.~F.;
  Burch,~K.~S.; Kee,~H.-Y.; Kim,~Y.-J. $\alpha$-{R}u{C}l$_3$: A spin-orbit
  assisted {M}ott insulator on a honeycomb lattice. \emph{Phys. Rev. B}
  \textbf{2014}, \emph{90}, 041112\relax
\mciteBstWouldAddEndPuncttrue
\mciteSetBstMidEndSepPunct{\mcitedefaultmidpunct}
{\mcitedefaultendpunct}{\mcitedefaultseppunct}\relax
\EndOfBibitem
\bibitem[Johnson \latin{et~al.}(2015)Johnson, Williams, Haghighirad, Singleton,
  Zapf, Manuel, Mazin, Li, Jeschke, Valent{\'\i}, and Coldea]{johnson_PRB2015}
Johnson,~R.~D.; Williams,~S.; Haghighirad,~A.; Singleton,~J.; Zapf,~V.;
  Manuel,~P.; Mazin,~I.; Li,~Y.; Jeschke,~H.~O.; Valent{\'\i},~R.; Coldea,~R.
  Monoclinic crystal structure of $\alpha$- RuCl 3 and the zigzag
  antiferromagnetic ground state. \emph{Physical Review B} \textbf{2015},
  \emph{92}, 235119\relax
\mciteBstWouldAddEndPuncttrue
\mciteSetBstMidEndSepPunct{\mcitedefaultmidpunct}
{\mcitedefaultendpunct}{\mcitedefaultseppunct}\relax
\EndOfBibitem
\bibitem[Rau \latin{et~al.}(2014)Rau, Lee, and Kee]{Rau_PRL2014}
Rau,~J.~G.; Lee,~E. K.-H.; Kee,~H.-Y. Generic Spin Model for the Honeycomb
  Iridates beyond the Kitaev Limit. \emph{Phys. Rev. Lett.} \textbf{2014},
  \emph{112}, 077204\relax
\mciteBstWouldAddEndPuncttrue
\mciteSetBstMidEndSepPunct{\mcitedefaultmidpunct}
{\mcitedefaultendpunct}{\mcitedefaultseppunct}\relax
\EndOfBibitem
\bibitem[Winter \latin{et~al.}(2016)Winter, Li, Jeschke, and
  Valent\'{\i}]{Winter_PRB2016}
Winter,~S.~M.; Li,~Y.; Jeschke,~H.~O.; Valent\'{\i},~R. Challenges in design of
  Kitaev materials: Magnetic interactions from competing energy scales.
  \emph{Phys. Rev. B} \textbf{2016}, \emph{93}, 214431\relax
\mciteBstWouldAddEndPuncttrue
\mciteSetBstMidEndSepPunct{\mcitedefaultmidpunct}
{\mcitedefaultendpunct}{\mcitedefaultseppunct}\relax
\EndOfBibitem
\bibitem[Winter \latin{et~al.}(2017)Winter, Riedl, Maksimov, Chernyshev,
  Honecker, and Valent{\'i}]{Winter_NatComm2017}
Winter,~S.~M.; Riedl,~K.; Maksimov,~P.~A.; Chernyshev,~A.~L.; Honecker,~A.;
  Valent{\'i},~R. Breakdown of magnons in a strongly spin-orbital coupled
  magnet. \emph{Nature Communications} \textbf{2017}, \emph{8}, 1152\relax
\mciteBstWouldAddEndPuncttrue
\mciteSetBstMidEndSepPunct{\mcitedefaultmidpunct}
{\mcitedefaultendpunct}{\mcitedefaultseppunct}\relax
\EndOfBibitem
\bibitem[Winter \latin{et~al.}(2017)Winter, Tsirlin, Daghofer, van~den Brink,
  Singh, Gegenwart, and Valent{\'\i}]{winter2017_review}
Winter,~S.~M.; Tsirlin,~A.~A.; Daghofer,~M.; van~den Brink,~J.; Singh,~Y.;
  Gegenwart,~P.; Valent{\'\i},~R. Models and materials for generalized Kitaev
  magnetism. \emph{Journal of Physics: Condensed Matter} \textbf{2017},
  \emph{29}, 493002\relax
\mciteBstWouldAddEndPuncttrue
\mciteSetBstMidEndSepPunct{\mcitedefaultmidpunct}
{\mcitedefaultendpunct}{\mcitedefaultseppunct}\relax
\EndOfBibitem
\bibitem[Maksimov and Chernyshev(2020)Maksimov, and
  Chernyshev]{Maksimov_PRR2020}
Maksimov,~P.~A.; Chernyshev,~A.~L. Rethinking
  $\ensuremath{\alpha}\text{\ensuremath{-}}{\mathrm{RuCl}}_{3}$. \emph{Phys.
  Rev. Res.} \textbf{2020}, \emph{2}, 033011\relax
\mciteBstWouldAddEndPuncttrue
\mciteSetBstMidEndSepPunct{\mcitedefaultmidpunct}
{\mcitedefaultendpunct}{\mcitedefaultseppunct}\relax
\EndOfBibitem
\bibitem[Rousochatzakis \latin{et~al.}(2023)Rousochatzakis, Perkins, Luo, and
  Kee]{rousochatzakis2023_review}
Rousochatzakis,~I.; Perkins,~N.~B.; Luo,~Q.; Kee,~H.-Y. Beyond Kitaev physics
  in strong spin-orbit coupled magnets. \emph{arXiv preprint arXiv:2308.01943}
  \textbf{2023}, \relax
\mciteBstWouldAddEndPunctfalse
\mciteSetBstMidEndSepPunct{\mcitedefaultmidpunct}
{}{\mcitedefaultseppunct}\relax
\EndOfBibitem
\bibitem[Yokoi \latin{et~al.}(2021)Yokoi, Ma, Kasahara, Kasahara, Shibauchi,
  Kurita, Tanaka, Nasu, Motome, Hickey, Trebst, and Matsuda]{Yokoi_Science2021}
Yokoi,~T.; Ma,~S.; Kasahara,~Y.; Kasahara,~S.; Shibauchi,~T.; Kurita,~N.;
  Tanaka,~H.; Nasu,~J.; Motome,~Y.; Hickey,~C.; Trebst,~S.; Matsuda,~Y.
  Half-integer quantized anomalous thermal Hall effect in the Kitaev material
  candidate $\ensuremath{\alpha}\ensuremath{-}{\mathrm{RuCl}}_{3}$.
  \emph{Science} \textbf{2021}, \emph{373}, 568--572\relax
\mciteBstWouldAddEndPuncttrue
\mciteSetBstMidEndSepPunct{\mcitedefaultmidpunct}
{\mcitedefaultendpunct}{\mcitedefaultseppunct}\relax
\EndOfBibitem
\bibitem[Lefran{\c{c}}ois \latin{et~al.}(2022)Lefran{\c{c}}ois, Grissonnanche,
  Baglo, Lampen-Kelley, Yan, Balz, Mandrus, Nagler, Kim, Kim, \latin{et~al.}
  others]{lefranccois2022}
Lefran{\c{c}}ois,~{\'E}.; Grissonnanche,~G.; Baglo,~J.; Lampen-Kelley,~P.;
  Yan,~J.-Q.; Balz,~C.; Mandrus,~D.; Nagler,~S.; Kim,~S.; Kim,~Y.-J.,
  \latin{et~al.}  Evidence of a Phonon Hall Effect in the Kitaev Spin Liquid
  Candidate $\alpha$- RuCl 3. \emph{Physical Review X} \textbf{2022},
  \emph{12}, 021025\relax
\mciteBstWouldAddEndPuncttrue
\mciteSetBstMidEndSepPunct{\mcitedefaultmidpunct}
{\mcitedefaultendpunct}{\mcitedefaultseppunct}\relax
\EndOfBibitem
\bibitem[Bruin \latin{et~al.}(2022)Bruin, Claus, Matsumoto, Kurita, Tanaka, and
  Takagi]{Bruin_NatPhys2022}
Bruin,~J. A.~N.; Claus,~R.~R.; Matsumoto,~Y.; Kurita,~N.; Tanaka,~H.;
  Takagi,~H. Robustness of the thermal Hall effect close to half-quantization
  in $\alpha$-RuCl3. \emph{Nature Physics} \textbf{2022}, \emph{18},
  401--405\relax
\mciteBstWouldAddEndPuncttrue
\mciteSetBstMidEndSepPunct{\mcitedefaultmidpunct}
{\mcitedefaultendpunct}{\mcitedefaultseppunct}\relax
\EndOfBibitem
\bibitem[Czajka \latin{et~al.}(2023)Czajka, Gao, Hirschberger, Lampen-Kelley,
  Banerjee, Quirk, Mandrus, Nagler, and Ong]{czajka_NatMat2023}
Czajka,~P.; Gao,~T.; Hirschberger,~M.; Lampen-Kelley,~P.; Banerjee,~A.;
  Quirk,~N.; Mandrus,~D.~G.; Nagler,~S.~E.; Ong,~N.~P. Planar thermal Hall
  effect of topological bosons in the Kitaev magnet $\alpha$-RuCl3.
  \emph{Nature Materials} \textbf{2023}, \emph{22}, 36--41\relax
\mciteBstWouldAddEndPuncttrue
\mciteSetBstMidEndSepPunct{\mcitedefaultmidpunct}
{\mcitedefaultendpunct}{\mcitedefaultseppunct}\relax
\EndOfBibitem
\bibitem[Bastien \latin{et~al.}(2019)Bastien, Roslova, Haghighi, Mehlawat,
  Hunger, Isaeva, Doert, Vojta, B{\"u}chner, and Wolter]{bastien_PRB2019}
Bastien,~G.; Roslova,~M.; Haghighi,~M.; Mehlawat,~K.; Hunger,~J.; Isaeva,~A.;
  Doert,~T.; Vojta,~M.; B{\"u}chner,~B.; Wolter,~A. Spin-glass state and
  reversed magnetic anisotropy induced by Cr doping in the Kitaev magnet
  $\alpha$- RuCl 3. \emph{Physical Review B} \textbf{2019}, \emph{99},
  214410\relax
\mciteBstWouldAddEndPuncttrue
\mciteSetBstMidEndSepPunct{\mcitedefaultmidpunct}
{\mcitedefaultendpunct}{\mcitedefaultseppunct}\relax
\EndOfBibitem
\bibitem[Bastien \latin{et~al.}(2018)Bastien, Garbarino, Yadav,
  Mart{\'\i}nez-Casado, Rodr{\'\i}guez, Stahl, Kusch, Limandri, Ray,
  Lampen-Kelley, \latin{et~al.} others]{bastien2018}
Bastien,~G.; Garbarino,~G.; Yadav,~R.; Mart{\'\i}nez-Casado,~F.~J.;
  Rodr{\'\i}guez,~R.~B.; Stahl,~Q.; Kusch,~M.; Limandri,~S.~P.; Ray,~R.;
  Lampen-Kelley,~P., \latin{et~al.}  Pressure-induced dimerization and valence
  bond crystal formation in the Kitaev-Heisenberg magnet $\alpha$- RuCl 3.
  \emph{Physical Review B} \textbf{2018}, \emph{97}, 241108\relax
\mciteBstWouldAddEndPuncttrue
\mciteSetBstMidEndSepPunct{\mcitedefaultmidpunct}
{\mcitedefaultendpunct}{\mcitedefaultseppunct}\relax
\EndOfBibitem
\bibitem[Bachus \latin{et~al.}(2020)Bachus, Kaib, Tokiwa, Jesche, Tsurkan,
  Loidl, Winter, Tsirlin, Valenti, and Gegenwart]{bachus2020}
Bachus,~S.; Kaib,~D.~A.; Tokiwa,~Y.; Jesche,~A.; Tsurkan,~V.; Loidl,~A.;
  Winter,~S.~M.; Tsirlin,~A.~A.; Valenti,~R.; Gegenwart,~P. Thermodynamic
  Perspective on Field-Induced Behavior of $\alpha$- RuCl 3. \emph{Physical
  Review Letters} \textbf{2020}, \emph{125}, 097203\relax
\mciteBstWouldAddEndPuncttrue
\mciteSetBstMidEndSepPunct{\mcitedefaultmidpunct}
{\mcitedefaultendpunct}{\mcitedefaultseppunct}\relax
\EndOfBibitem
\bibitem[Kocsis \latin{et~al.}(2022)Kocsis, Kaib, Riedl, Gass, Lampen-Kelley,
  Mandrus, Nagler, P{\'e}rez, Nielsch, B{\"u}chner, \latin{et~al.}
  others]{kocsis2022}
Kocsis,~V.; Kaib,~D.~A.; Riedl,~K.; Gass,~S.; Lampen-Kelley,~P.;
  Mandrus,~D.~G.; Nagler,~S.~E.; P{\'e}rez,~N.; Nielsch,~K.; B{\"u}chner,~B.,
  \latin{et~al.}  Magnetoelastic coupling anisotropy in the Kitaev material
  $\alpha$- Ru Cl 3. \emph{Physical Review B} \textbf{2022}, \emph{105},
  094410\relax
\mciteBstWouldAddEndPuncttrue
\mciteSetBstMidEndSepPunct{\mcitedefaultmidpunct}
{\mcitedefaultendpunct}{\mcitedefaultseppunct}\relax
\EndOfBibitem
\bibitem[Kaib \latin{et~al.}(2021)Kaib, Biswas, Riedl, Winter, and
  Valent{\'\i}]{kaib2021}
Kaib,~D.~A.; Biswas,~S.; Riedl,~K.; Winter,~S.~M.; Valent{\'\i},~R.
  Magnetoelastic coupling and effects of uniaxial strain in $\alpha$- RuCl 3
  from first principles. \emph{Physical Review B} \textbf{2021}, \emph{103},
  L140402\relax
\mciteBstWouldAddEndPuncttrue
\mciteSetBstMidEndSepPunct{\mcitedefaultmidpunct}
{\mcitedefaultendpunct}{\mcitedefaultseppunct}\relax
\EndOfBibitem
\bibitem[Mashhadi \latin{et~al.}(2019)Mashhadi, Kim, Kim, Weber, Taniguchi,
  Watanabe, Park, Lotsch, Smet, Burghard, \latin{et~al.} others]{mashhadi2019}
Mashhadi,~S.; Kim,~Y.; Kim,~J.; Weber,~D.; Taniguchi,~T.; Watanabe,~K.;
  Park,~N.; Lotsch,~B.; Smet,~J.~H.; Burghard,~M., \latin{et~al.}  Spin-split
  band hybridization in graphene proximitized with $\alpha$-RuCl3 nanosheets.
  \emph{Nano letters} \textbf{2019}, \emph{19}, 4659--4665\relax
\mciteBstWouldAddEndPuncttrue
\mciteSetBstMidEndSepPunct{\mcitedefaultmidpunct}
{\mcitedefaultendpunct}{\mcitedefaultseppunct}\relax
\EndOfBibitem
\bibitem[Zhou \latin{et~al.}(2019)Zhou, Balgley, Lampen-Kelley, Yan, Mandrus,
  and Henriksen]{zhou2019}
Zhou,~B.; Balgley,~J.; Lampen-Kelley,~P.; Yan,~J.-Q.; Mandrus,~D.~G.;
  Henriksen,~E.~A. Evidence for charge transfer and proximate magnetism in
  graphene--$\alpha$- RuCl 3 heterostructures. \emph{Physical Review B}
  \textbf{2019}, \emph{100}, 165426\relax
\mciteBstWouldAddEndPuncttrue
\mciteSetBstMidEndSepPunct{\mcitedefaultmidpunct}
{\mcitedefaultendpunct}{\mcitedefaultseppunct}\relax
\EndOfBibitem
\bibitem[Biswas \latin{et~al.}(2019)Biswas, Li, Winter, Knolle, and
  Valent{\'\i}]{biswas2019}
Biswas,~S.; Li,~Y.; Winter,~S.~M.; Knolle,~J.; Valent{\'\i},~R. Electronic
  Properties of $\alpha$- RuCl 3 in proximity to graphene. \emph{Physical
  Review Letters} \textbf{2019}, \emph{123}, 237201\relax
\mciteBstWouldAddEndPuncttrue
\mciteSetBstMidEndSepPunct{\mcitedefaultmidpunct}
{\mcitedefaultendpunct}{\mcitedefaultseppunct}\relax
\EndOfBibitem
\bibitem[Rizzo \latin{et~al.}(2020)Rizzo, Jessen, Sun, Ruta, Zhang, Yan, Xian,
  McLeod, Berkowitz, Watanabe, Taniguchi, Nagler, Mandrus, Rubio, Fogler,
  Millis, Hone, Dean, and Basov]{Rizzo_NanoLett2020}
Rizzo,~D.~J. \latin{et~al.}  Charge-Transfer Plasmon Polaritons at
  Graphene/$\alpha$-RuCl3 Interfaces. \emph{Nano Letters} \textbf{2020},
  \emph{20}, 8438--8445\relax
\mciteBstWouldAddEndPuncttrue
\mciteSetBstMidEndSepPunct{\mcitedefaultmidpunct}
{\mcitedefaultendpunct}{\mcitedefaultseppunct}\relax
\EndOfBibitem
\bibitem[Gerber \latin{et~al.}(2020)Gerber, Yao, Arias, and Kim]{gerber2020}
Gerber,~E.; Yao,~Y.; Arias,~T.~A.; Kim,~E.-A. Ab Initio Mismatched Interface
  Theory of Graphene on $\alpha$- RuCl 3: Doping and Magnetism. \emph{Physical
  Review Letters} \textbf{2020}, \emph{124}, 106804\relax
\mciteBstWouldAddEndPuncttrue
\mciteSetBstMidEndSepPunct{\mcitedefaultmidpunct}
{\mcitedefaultendpunct}{\mcitedefaultseppunct}\relax
\EndOfBibitem
\bibitem[Leeb \latin{et~al.}(2021)Leeb, Polyudov, Mashhadi, Biswas,
  Valent{\'\i}, Burghard, and Knolle]{leeb_PRL2021}
Leeb,~V.; Polyudov,~K.; Mashhadi,~S.; Biswas,~S.; Valent{\'\i},~R.;
  Burghard,~M.; Knolle,~J. Anomalous quantum oscillations in a heterostructure
  of graphene on a proximate quantum spin liquid. \emph{Physical Review
  Letters} \textbf{2021}, \emph{126}, 097201\relax
\mciteBstWouldAddEndPuncttrue
\mciteSetBstMidEndSepPunct{\mcitedefaultmidpunct}
{\mcitedefaultendpunct}{\mcitedefaultseppunct}\relax
\EndOfBibitem
\bibitem[Balgley \latin{et~al.}(2022)Balgley, Butler, Biswas, Ge, Lagasse,
  Taniguchi, Watanabe, Cothrine, Mandrus, Velasco~Jr, \latin{et~al.}
  others]{balgley2022}
Balgley,~J.; Butler,~J.; Biswas,~S.; Ge,~Z.; Lagasse,~S.; Taniguchi,~T.;
  Watanabe,~K.; Cothrine,~M.; Mandrus,~D.~G.; Velasco~Jr,~J., \latin{et~al.}
  Ultrasharp Lateral p--n Junctions in Modulation-Doped Graphene. \emph{Nano
  Letters} \textbf{2022}, \emph{22}, 4124--4130\relax
\mciteBstWouldAddEndPuncttrue
\mciteSetBstMidEndSepPunct{\mcitedefaultmidpunct}
{\mcitedefaultendpunct}{\mcitedefaultseppunct}\relax
\EndOfBibitem
\bibitem[Shi and MacDonald(2023)Shi, and MacDonald]{shi2023magnetic}
Shi,~J.; MacDonald,~A. Magnetic states of graphene proximitized Kitaev
  materials. \emph{Physical Review B} \textbf{2023}, \emph{108}, 064401\relax
\mciteBstWouldAddEndPuncttrue
\mciteSetBstMidEndSepPunct{\mcitedefaultmidpunct}
{\mcitedefaultendpunct}{\mcitedefaultseppunct}\relax
\EndOfBibitem
\bibitem[Cao \latin{et~al.}(2018)Cao, Fatemi, Fang, Watanabe, Taniguchi,
  Kaxiras, and Jarillo-Herrero]{Cao2018}
Cao,~Y.; Fatemi,~V.; Fang,~S.; Watanabe,~K.; Taniguchi,~T.; Kaxiras,~E.;
  Jarillo-Herrero,~P. Unconventional superconductivity in magic-angle graphene
  superlattices. \emph{Nature} \textbf{2018}, \emph{556}, 43--50\relax
\mciteBstWouldAddEndPuncttrue
\mciteSetBstMidEndSepPunct{\mcitedefaultmidpunct}
{\mcitedefaultendpunct}{\mcitedefaultseppunct}\relax
\EndOfBibitem
\bibitem[Devakul \latin{et~al.}(2021)Devakul, Cr{\'e}pel, Zhang, and
  Fu]{Devakul_NatComm2021}
Devakul,~T.; Cr{\'e}pel,~V.; Zhang,~Y.; Fu,~L. Magic in twisted transition
  metal dichalcogenide bilayers. \emph{Nature Communications} \textbf{2021},
  \emph{12}, 6730\relax
\mciteBstWouldAddEndPuncttrue
\mciteSetBstMidEndSepPunct{\mcitedefaultmidpunct}
{\mcitedefaultendpunct}{\mcitedefaultseppunct}\relax
\EndOfBibitem
\bibitem[Zhao \latin{et~al.}(2021)Zhao, Poccia, Cui, Volkov, Yoo, Engelke,
  Ronen, Zhong, Gu, Plugge, Tummuru, Franz, Pixley, and Kim]{Zhao_arxiv2021}
Zhao,~S. Y.~F.; Poccia,~N.; Cui,~X.; Volkov,~P.~A.; Yoo,~H.; Engelke,~R.;
  Ronen,~Y.; Zhong,~R.; Gu,~G.; Plugge,~S.; Tummuru,~T.; Franz,~M.;
  Pixley,~J.~H.; Kim,~P. Emergent Interfacial Superconductivity between Twisted
  Cuprate Superconductors. \emph{arXiv e-prints} \textbf{2021}, \relax
\mciteBstWouldAddEndPunctfalse
\mciteSetBstMidEndSepPunct{\mcitedefaultmidpunct}
{}{\mcitedefaultseppunct}\relax
\EndOfBibitem
\bibitem[Tong \latin{et~al.}(2018)Tong, Liu, Xiao, and Yao]{Tong_ACS2018}
Tong,~Q.; Liu,~F.; Xiao,~J.; Yao,~W. Skyrmions in the moir{\'e} of van der
  {Waals} 2D Magnets. \emph{Nano Letters} \textbf{2018}, \emph{18},
  7194--7199\relax
\mciteBstWouldAddEndPuncttrue
\mciteSetBstMidEndSepPunct{\mcitedefaultmidpunct}
{\mcitedefaultendpunct}{\mcitedefaultseppunct}\relax
\EndOfBibitem
\bibitem[Hejazi \latin{et~al.}(2020)Hejazi, Luo, and Balents]{Hejazi_PNAS2020}
Hejazi,~K.; Luo,~Z.-X.; Balents,~L. Noncollinear phases in moir{\'e} magnets.
  \emph{Proceedings of the National Academy of Sciences} \textbf{2020},
  \emph{117}, 10721--10726\relax
\mciteBstWouldAddEndPuncttrue
\mciteSetBstMidEndSepPunct{\mcitedefaultmidpunct}
{\mcitedefaultendpunct}{\mcitedefaultseppunct}\relax
\EndOfBibitem
\bibitem[Hejazi \latin{et~al.}(2021)Hejazi, Luo, and Balents]{Hejazi_PRB2021}
Hejazi,~K.; Luo,~Z.-X.; Balents,~L. Heterobilayer moir\'e magnets: Moir\'e
  skyrmions and commensurate-incommensurate transitions. \emph{Phys. Rev. B}
  \textbf{2021}, \emph{104}, L100406\relax
\mciteBstWouldAddEndPuncttrue
\mciteSetBstMidEndSepPunct{\mcitedefaultmidpunct}
{\mcitedefaultendpunct}{\mcitedefaultseppunct}\relax
\EndOfBibitem
\bibitem[Akram and Erten(2021)Akram, and Erten]{Akram_PRB2021}
Akram,~M.; Erten,~O. Skyrmions in twisted van der Waals magnets. \emph{Phys.
  Rev. B} \textbf{2021}, \emph{103}, L140406\relax
\mciteBstWouldAddEndPuncttrue
\mciteSetBstMidEndSepPunct{\mcitedefaultmidpunct}
{\mcitedefaultendpunct}{\mcitedefaultseppunct}\relax
\EndOfBibitem
\bibitem[Akram \latin{et~al.}(2021)Akram, LaBollita, Dey, Kapeghian, Erten, and
  Botana]{Akram_NanoLett2021}
Akram,~M.; LaBollita,~H.; Dey,~D.; Kapeghian,~J.; Erten,~O.; Botana,~A.~S.
  Moir{\'e} Skyrmions and Chiral Magnetic Phases in Twisted CrX3 (X = I, Br,
  and Cl) Bilayers. \emph{Nano Letters} \textbf{2021}, \emph{21},
  6633--6639\relax
\mciteBstWouldAddEndPuncttrue
\mciteSetBstMidEndSepPunct{\mcitedefaultmidpunct}
{\mcitedefaultendpunct}{\mcitedefaultseppunct}\relax
\EndOfBibitem
\bibitem[Xiao \latin{et~al.}(2021)Xiao, Chen, and Tong]{Tong_PRR2021}
Xiao,~F.; Chen,~K.; Tong,~Q. Magnetization textures in twisted bilayer
  $\mathrm{Cr}\mathrm{X}_{3}$ ($\mathrm{X} = \mathrm{Br}, \mathrm{I})$.
  \emph{Phys. Rev. Research} \textbf{2021}, \emph{3}, 013027\relax
\mciteBstWouldAddEndPuncttrue
\mciteSetBstMidEndSepPunct{\mcitedefaultmidpunct}
{\mcitedefaultendpunct}{\mcitedefaultseppunct}\relax
\EndOfBibitem
\bibitem[Xiao \latin{et~al.}(2021)Xiao, Chen, and Tong]{Xiao_PRR2021}
Xiao,~F.; Chen,~K.; Tong,~Q. Magnetization textures in twisted bilayer
  $\mathrm{Cr}{\mathrm{X}}_{3}$ ($\mathrm{X}$=Br, I). \emph{Phys. Rev. Res.}
  \textbf{2021}, \emph{3}, 013027\relax
\mciteBstWouldAddEndPuncttrue
\mciteSetBstMidEndSepPunct{\mcitedefaultmidpunct}
{\mcitedefaultendpunct}{\mcitedefaultseppunct}\relax
\EndOfBibitem
\bibitem[Ghader \latin{et~al.}(2022)Ghader, Jabakhanji, and
  Stroppa]{Ghader_CommPhys2022}
Ghader,~D.; Jabakhanji,~B.; Stroppa,~A. Whirling interlayer fields as a source
  of stable topological order in moir{\'e} CrI3. \emph{Communications Physics}
  \textbf{2022}, \emph{5}, 192\relax
\mciteBstWouldAddEndPuncttrue
\mciteSetBstMidEndSepPunct{\mcitedefaultmidpunct}
{\mcitedefaultendpunct}{\mcitedefaultseppunct}\relax
\EndOfBibitem
\bibitem[Zheng(2023)]{Zheng_AFM2023}
Zheng,~F. Magnetic Skyrmion Lattices in a Novel 2D-Twisted Bilayer Magnet.
  \emph{Advanced Functional Materials} \textbf{2023}, \emph{33}, 2206923\relax
\mciteBstWouldAddEndPuncttrue
\mciteSetBstMidEndSepPunct{\mcitedefaultmidpunct}
{\mcitedefaultendpunct}{\mcitedefaultseppunct}\relax
\EndOfBibitem
\bibitem[Kim \latin{et~al.}(2023)Kim, Kiem, Bednik, Han, and
  Park]{Kim_NanoLett2023}
Kim,~K.-M.; Kiem,~D.~H.; Bednik,~G.; Han,~M.~J.; Park,~M.~J. Ab Initio Spin
  Hamiltonian and Topological Noncentrosymmetric Magnetism in Twisted Bilayer
  CrI3. \emph{Nano Letters} \textbf{2023}, \emph{23}, 6088--6094\relax
\mciteBstWouldAddEndPuncttrue
\mciteSetBstMidEndSepPunct{\mcitedefaultmidpunct}
{\mcitedefaultendpunct}{\mcitedefaultseppunct}\relax
\EndOfBibitem
\bibitem[Fumega and Lado(2023)Fumega, and Lado]{Fumega_2DMAtt2023}
Fumega,~A.~O.; Lado,~J.~L. Moir\'e-driven multiferroic order in twisted CrCl3,
  CrBr3 and CrI3 bilayers. \emph{2D Materials} \textbf{2023}, \emph{10},
  025026\relax
\mciteBstWouldAddEndPuncttrue
\mciteSetBstMidEndSepPunct{\mcitedefaultmidpunct}
{\mcitedefaultendpunct}{\mcitedefaultseppunct}\relax
\EndOfBibitem
\bibitem[Xu \latin{et~al.}(2021)Xu, Ray, Shao, Jiang, Lee, Weber, Goldberger,
  Watanabe, Taniguchi, Muller, Mak, and Shan]{Xu_NatNano2021}
Xu,~Y.; Ray,~A.; Shao,~Y.-T.; Jiang,~S.; Lee,~K.; Weber,~D.; Goldberger,~J.~E.;
  Watanabe,~K.; Taniguchi,~T.; Muller,~D.~A.; Mak,~K.~F.; Shan,~J. Coexisting
  ferromagnetic--antiferromagnetic state in twisted bilayer CrI3. \emph{Nature
  Nanotechnology} \textbf{2021}, \relax
\mciteBstWouldAddEndPunctfalse
\mciteSetBstMidEndSepPunct{\mcitedefaultmidpunct}
{}{\mcitedefaultseppunct}\relax
\EndOfBibitem
\bibitem[Song \latin{et~al.}(2021)Song, Sun, Anderson, Wang, Qian, Taniguchi,
  Watanabe, McGuire, Stšhr, Xiao, Cao, Wrachtrup, and Xu]{Song_Science2021}
Song,~T.; Sun,~Q.-C.; Anderson,~E.; Wang,~C.; Qian,~J.; Taniguchi,~T.;
  Watanabe,~K.; McGuire,~M.~A.; St\"ohr,~R.; Xiao,~D.; Cao,~T.; Wrachtrup,~J.;
  Xu,~X. Direct visualization of magnetic domains and moire magnetism in
  twisted 2D magnets. \emph{Science} \textbf{2021}, \emph{374},
  1140--1144\relax
\mciteBstWouldAddEndPuncttrue
\mciteSetBstMidEndSepPunct{\mcitedefaultmidpunct}
{\mcitedefaultendpunct}{\mcitedefaultseppunct}\relax
\EndOfBibitem
\bibitem[Xie \latin{et~al.}(2023)Xie, Luo, Ye, Sun, Ye, Sung, Ge, Yan, Fu,
  Tian, Lei, Sun, Hovden, He, and Zhao]{xie2022}
Xie,~H.; Luo,~X.; Ye,~Z.; Sun,~Z.; Ye,~G.; Sung,~S.~H.; Ge,~H.; Yan,~S.;
  Fu,~Y.; Tian,~S.; Lei,~H.; Sun,~K.; Hovden,~R.; He,~R.; Zhao,~L. Evidence of
  non-collinear spin texture in magnetic moir{\'e} superlattices. \emph{Nature
  Physics} \textbf{2023}, \relax
\mciteBstWouldAddEndPunctfalse
\mciteSetBstMidEndSepPunct{\mcitedefaultmidpunct}
{}{\mcitedefaultseppunct}\relax
\EndOfBibitem
\bibitem[Lee \latin{et~al.}(2021)Lee, Choi, Do, Kim, Seong, and
  Choi]{Lee_npjQM2021}
Lee,~J.-H.; Choi,~Y.; Do,~S.-H.; Kim,~B.~H.; Seong,~M.-J.; Choi,~K.-Y. Multiple
  spin-orbit excitons in $\alpha$-RuCl3 from bulk to atomically thin layers.
  \emph{npj Quantum Materials} \textbf{2021}, \emph{6}, 43\relax
\mciteBstWouldAddEndPuncttrue
\mciteSetBstMidEndSepPunct{\mcitedefaultmidpunct}
{\mcitedefaultendpunct}{\mcitedefaultseppunct}\relax
\EndOfBibitem
\bibitem[Yang \latin{et~al.}(2023)Yang, Goh, Sung, Ye, Biswas, Kaib, Dhakal,
  Yan, Li, Jiang, Chen, Lei, He, Valent{\'i}, Winter, Hovden, and
  Tsen]{Yang_NatMat2023}
Yang,~B. \latin{et~al.}  Magnetic anisotropy reversal driven by structural
  symmetry-breaking in monolayer $\alpha$-RuCl3. \emph{Nature Materials}
  \textbf{2023}, \emph{22}, 50--57\relax
\mciteBstWouldAddEndPuncttrue
\mciteSetBstMidEndSepPunct{\mcitedefaultmidpunct}
{\mcitedefaultendpunct}{\mcitedefaultseppunct}\relax
\EndOfBibitem
\bibitem[Janssen \latin{et~al.}(2020)Janssen, Koch, and Vojta]{Janssen_RPB2020}
Janssen,~L.; Koch,~S.; Vojta,~M. Magnon dispersion and dynamic spin response in
  three-dimensional spin models for
  $\ensuremath{\alpha}\text{\ensuremath{-}}{\mathrm{RuCl}}_{3}$. \emph{Phys.
  Rev. B} \textbf{2020}, \emph{101}, 174444\relax
\mciteBstWouldAddEndPuncttrue
\mciteSetBstMidEndSepPunct{\mcitedefaultmidpunct}
{\mcitedefaultendpunct}{\mcitedefaultseppunct}\relax
\EndOfBibitem
\bibitem[Balz \latin{et~al.}(2021)Balz, Janssen, Lampen-Kelley, Banerjee, Liu,
  Yan, Mandrus, Vojta, and Nagler]{Balz_PRB2021}
Balz,~C.; Janssen,~L.; Lampen-Kelley,~P.; Banerjee,~A.; Liu,~Y.~H.; Yan,~J.-Q.;
  Mandrus,~D.~G.; Vojta,~M.; Nagler,~S.~E. Field-induced intermediate ordered
  phase and anisotropic interlayer interactions in
  $\ensuremath{\alpha}\text{\ensuremath{-}}{\mathrm{RuCl}}_{3}$. \emph{Phys.
  Rev. B} \textbf{2021}, \emph{103}, 174417\relax
\mciteBstWouldAddEndPuncttrue
\mciteSetBstMidEndSepPunct{\mcitedefaultmidpunct}
{\mcitedefaultendpunct}{\mcitedefaultseppunct}\relax
\EndOfBibitem
\bibitem[{Gilbert}(2004)]{1353448}
{Gilbert},~T.~L. A phenomenological theory of damping in ferromagnetic
  materials. \emph{IEEE Transactions on Magnetics} \textbf{2004}, \emph{40},
  3443--3449\relax
\mciteBstWouldAddEndPuncttrue
\mciteSetBstMidEndSepPunct{\mcitedefaultmidpunct}
{\mcitedefaultendpunct}{\mcitedefaultseppunct}\relax
\EndOfBibitem
\end{mcitethebibliography}
\providecommand{\latin}[1]{#1}
\makeatletter
\providecommand{\doi}
  {\begingroup\let\do\@makeother\dospecials
  \catcode`\{=1 \catcode`\}=2 \doi@aux}
\providecommand{\doi@aux}[1]{\endgroup\texttt{#1}}
\makeatother
\providecommand*\mcitethebibliography{\thebibliography}
\csname @ifundefined\endcsname{endmcitethebibliography}
  {\let\endmcitethebibliography\endthebibliography}{}

\end{document}


\title{Supporting Information for ``Theory of moir\'e magnetism in twisted bilayer $\alpha$-RuCl$_3$''}

\author{Muhammad Akram}
\affiliation{Department of Physics, Arizona State University, Tempe, AZ 85287, USA}
\author{Jesse Kapeghian}
\affiliation{Department of Physics, Arizona State University, Tempe, AZ 85287, USA}
\author{Jyotirish Das}
\affiliation{Department of Physics, Arizona State University, Tempe, AZ 85287, USA}
\author{Antia S. Botana}
\affiliation{Department of Physics, Arizona State University, Tempe, AZ 85287, USA}
\email{antia.botana@asu.edu}
\author{Onur Erten}
\affiliation{Department of Physics, Arizona State University, Tempe, AZ 85287, USA}
\email{onur.erten@asu.edu}

\date{\today}

\maketitle

\section{\label{sec:details}Technical details of the DFT  calculations}

We performed density-functional theory (DFT)-based calculations using projector augmented wave (PAW) pseudopotentials \cite{Kresse_PRB_1999} as implemented in the VASP code \cite{Kresse_PRB_1996,Kresse_CMS_1996}. For the exchange-correlation functional, the Perdew-Burke-Ernzerhof (PBE) \cite{Perdew_PRL_1996} version of the generalized gradient approximation (GGA) was chosen, on top of which the DFT-D3 van der Waals correction scheme \cite{Grimme_JCP_2010} was added for structural relaxations. In agreement with the literature, an on-site Coulomb repulsion $U$ was included to account for correlation effects in the Ru-$d$ electrons \cite{Rohrbach_JPCM_2003}, where the Liechtenstein \cite{Liechtenstein_PRB_1995} approach was used for the double-counting correction. The particular Hubbard $U$ value used (1.5 eV) was chosen by comparing the direct charge gap with the measured optical gap in the literature \cite{Plumb_PRB_2014} and the Hund's coupling $J_{H}$ value (0.3 eV) was chosen to be about 20\% of $U$. Spin-orbit coupling (SOC) was also included in the calculations. The wave functions were expanded in the plane-wave basis with a kinetic-energy cut-off of 450 eV. The 4d and 5s orbitals (4d$^{7}$5s$^{1}$ configuration) were considered as valence states for the Ru atoms while for the Cl atoms the 3s and 3p orbitals (3s$^{2}$3p$^{5}$ configuration) were considered as valence. Sampling over the Brillouin zone (BZ) was performed with a $9 \times 5 \times 2$ Monkhorst-Pack $k$-mesh centered on $\Gamma$.

\begin{figure}[H]
\begin{center}
\includegraphics[width=0.4\columnwidth]{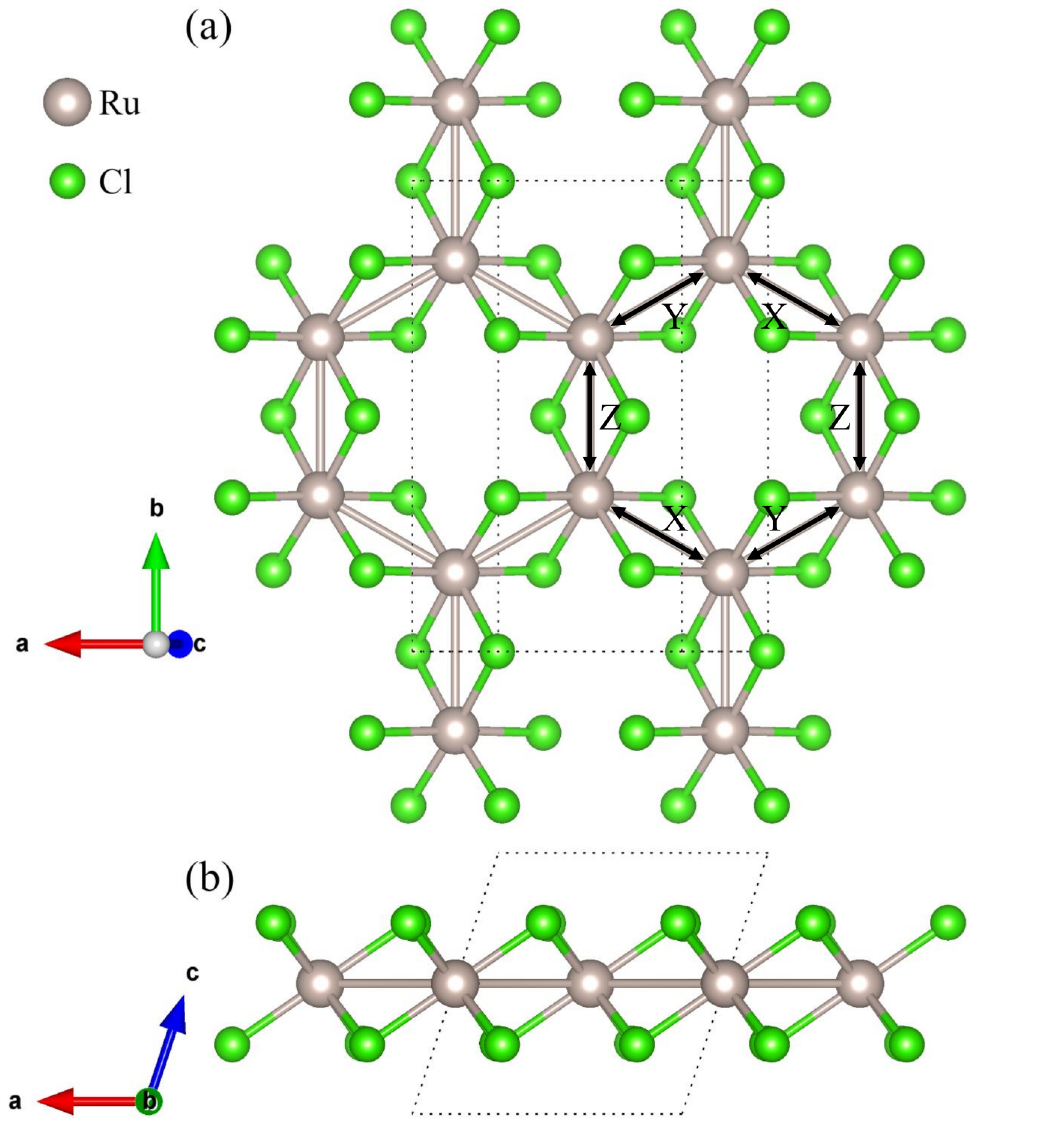}
\caption{Crystal structure of bulk RuCl$_{3}$ in the monoclinic ($C2/m$) phase showing the in-plane (a) and out-of-plane (b) arrangement of atoms where the larger gray spheres represent Ru atoms, the smaller green spheres represent Cl atoms, and the unit cell boundary is marked by a black dotted line. The three inequivalent nearest neighbor in-plane Ru-Ru bonds (X, Y, and Z) are indicated in (a).}
\label{fig:s1}
\end{center}
\end{figure}

We start by relaxing the monoclinic bulk structure (as pictured in Fig. \ref{fig:s1}) using the parameters described in the previous paragraph. We obtained in-plane lattice parameter magnitudes of $a=5.974$ \AA \, and $b=10.410$ \AA \, (consistent with the experimental values of 5.981 and 10.354 \AA, respectively \cite{Cao_PRB_2016}), and with angles $\alpha = \gamma = 90.000^{\circ}$ and $\beta = 108.280^{\circ}$ (the latter being slightly smaller than the experimental value of 108.800$^{\circ}$ \cite{Cao_PRB_2016}). Note in Fig. \ref{fig:s1} the three inequivalent bond types between nearest neighbor in-plane Ru atoms: X, Y, and Z. The obtained bond lengths are $\mathrm{X}=\mathrm{Y}=3.446$ \AA \, (consistent with the experimental value of 3.454 \AA \cite{Cao_PRB_2016}) and $\mathrm{Z}=3.487$ \AA \, (slightly larger than the experimental value of 3.448 \AA \cite{Cao_PRB_2016}). 

\begin{figure*}
\includegraphics[width=\columnwidth]{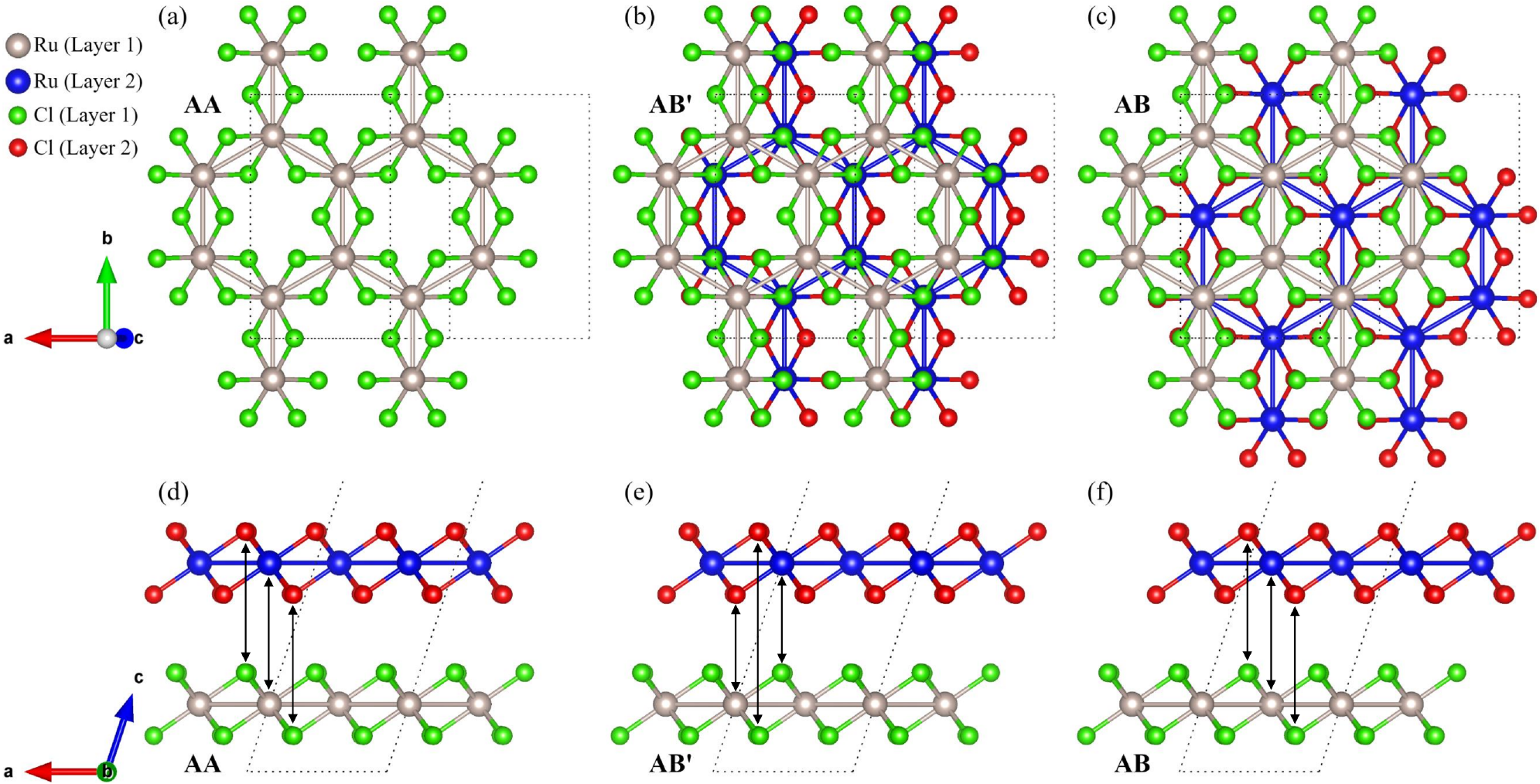}
\caption{Different stacking sequences for bilayer RuCl$_{3}$. (a-c) Top view showing the in-plane arrangement of atoms for AA, AB', and AB stackings. (d-f) Side view showing out-of-plane arrangement of atoms for AA, AB', and AB stackings. Solid black arrows indicate the alignment of atoms between layers: AA and AB stackings have Ru atoms on top of Ru atoms and Cl atoms on top of Cl atoms (but note that in AB, half the Ru atoms are aligned with gaps in hexagons of the other layer) while AB' stacking has Ru atoms aligned with Cl atoms. The larger gray (blue) spheres represent Ru atoms in the first (second) layer and the smaller green (red) spheres represent Cl atoms in the first (second) layer. In each figure the unit cell boundary is marked by a black dotted line.}
\label{fig:s2}
\end{figure*}

The bilayers were built from this relaxed bulk structure. Fig. \ref{fig:s2} shows the three bilayer stackings considered: AA, AB', and AB. The AA stacking (corresponding to a $P \overline{3} 1m$ space group) has the two layers directly aligned when viewed out-of-plane along the $c^{*}$ (i.e. $\hat{z}$) axis. AB' stacking ($C2/m$) is attained by a shift of the top layer by 1/3 along the $\left[ \, \overline{1} \, 0 \, 0 \, \right]$ direction (with respect to AA stacking), resulting in Ru atoms aligned with Cl atoms. Lastly, AB stacking ($R \overline{3} m$) is reached from AB' via a translation of the top layer by 1/6 along the $\left[ \, \overline{1} \, \overline{1} \, 0 \, \right]$ direction, leading to Cl atoms on top of Cl atoms and half of the Ru atoms aligned with Ru atoms along $c^{*}$ while the other half fall into the "gaps" or "hollow" sites in the hexagons of the adjacent layer. The interlayer spacing for the bilayers was determined by the distance required to shift the top layer by $a/3$ with respect to the bottom layer (as needed for AB' stacking order). Since $\tan \theta = (a/3) / d_{\mathrm{inter}}$, where $\theta=\beta-\pi/2$, one obtains $d_{\mathrm{inter}}=(a/3) / \tan \theta \sim 6.03$ \AA. The out-of-plane lattice parameter was then chosen to be $c=27.14$ \AA \, ($c_{z}=25.77$ \AA), with a vacuum of roughly 20 \AA \, along ${c}$ in order to suppress out-of-plane interactions between neighboring bilayers.

As mentioned in the main text, in order to determine $J^{\perp}$ we use the energy difference between FM and AFM (FM planes coupled AFM out of plane) spin configurations ($\Delta$E) obtained from DFT calculations performed for the three bilayer stackings (AA, AB', and AB). Using the notation of Eqs 4-6 in the main text, in the AA stacking, the following energy equations are obtained for the two spin states: $E^{\mathrm{FM}}_{\mathrm{AA}} = E_{0} + 4J_\perp(0)+12J_\perp(a_0)$ and $E^{\mathrm{AFM}}_{\mathrm{AA}} = E_{0} - 4J_\perp(0)- 12J_\perp(a_0)$. Here, $E_{0}$ is the total energy for the system omitting magnetic interactions. Next, for AB' stacking  $E^{\mathrm{FM}}_{\mathrm{AB'}} = E_{0} + 8 J_\perp(a_0/3)+8J_\perp(2a_0/3) $ and $E^{\mathrm{AFM}}_{\mathrm{AB'}} = E_{0} - 8 J_\perp(a_0/3)- 8J_\perp(2a_0/3) $. Lastly, for AB stacking $E^{\mathrm{FM}}_{\mathrm{AB}} = E_{0} + 2 J_\perp(0)+18 J_\perp(a_0)$ and $E^{\mathrm{AFM}}_{\mathrm{AB}} = E_{0} - 2 J_\perp(0)- 18 J_\perp(a_0) $. Note that in each of the expressions above, the energy maps correspond to one unit cell (or eight formula units). Using the energy difference between these two magnetic configurations, the interlayer exchange was derived as shown in the main text. For all stackings, the interlayer exchange is AFM.  

\subsection{Estimation of domain wall energy cost}
\begin{figure}
\center
\includegraphics[width=0.5\textwidth]{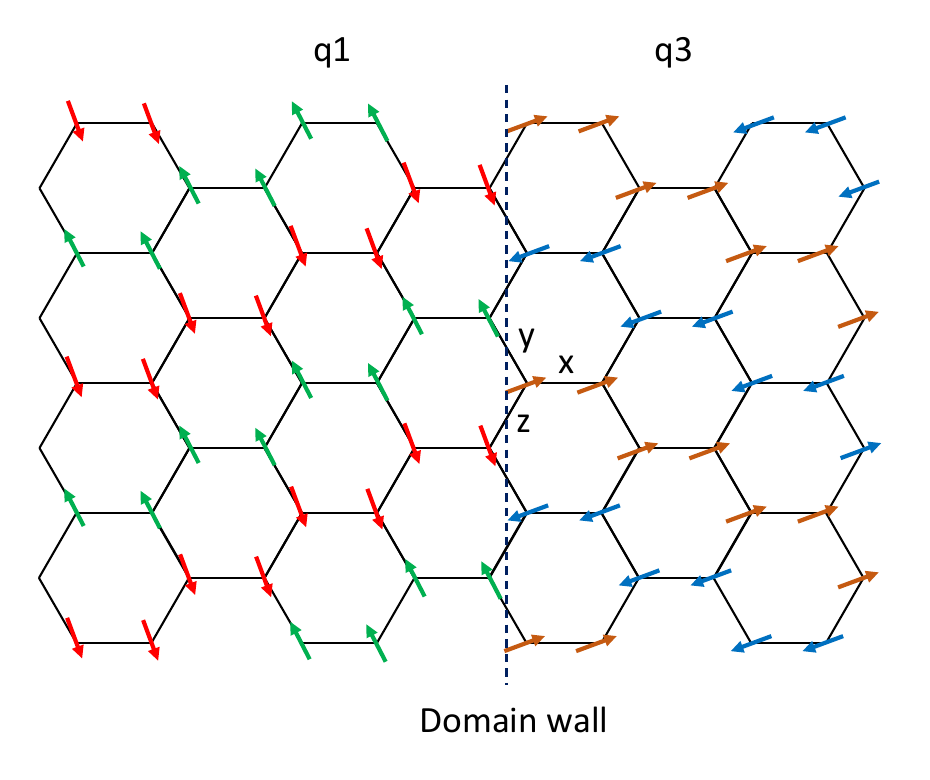} 
\caption{Schematic of a sharp domain wall between $q_1$ and $q_3$ zigzag orders.}
\label{domain_wall}
\end{figure}

In order to determine the energy cost of a domain wall between two different zigzag orders, we consider a sharp domain wall which separates $q_1$ and $q_3$ domains as shown in Fig.~\ref{domain_wall}. The direction of red spins in $q_1$ zigzag order is $s_1=(0.68,-0.25,0.68)$ (see Fig.~\ref{domain_wall}). The green spins are anti-parallel to red spin ($-s_1$). Similarly, the brown spins in $q_3$ domain point towards $s_3=(0.68,0.68,-0.25)$ and the blue spins are anti-parallel to brown spins ($-s_3$). At the domain wall the nearest neighbours of $s_3$ changes to $-s_1$ along y-bonds to $s_1$ along z-bonds. The energy along the y-bond in $q_3$ zigzag order is E$_{y-bond}=s_3J_y s_3\simeq-3.7084$ meV where $J_y=\begin{pmatrix}
J_1 & 0 & \Gamma\\
0 & J_1+K& 0\\
\Gamma & 0 & J_1
\end{pmatrix}$ is the exchange matrix along y-bond.
Similarly the energy along z-bond it is E$_{z-bond}=s_3J_z (-s_3)\simeq-1.5140$ meV where $J_z=\begin{pmatrix}
J_1 & \Gamma & 0\\
\Gamma & J_1& 0\\
0 & 0 & J_1+K
\end{pmatrix}$ is the exchange matrix along z-bond. 
The total energy along y and z-bonds is $\simeq-5.2223$ meV. The energies at domain wall along y and z-bonds are  E$_{y-bond}^{DW}=s_3J_y (-s_1)\simeq-1.5445$ meV and E$_{z-bond}^{DW}=s_3J_z s_1\simeq1.5445$ meV respectively and the total energy along these two bonds is zero. Therefore, the domain wall cost along these two bonds is $\simeq5.2223$ meV. Moreover, the third nearest bond of $s_3$ changes from $-s_3$ to $-s_1$ along horizontal axis. The energy due to third nearest neighbour in $q_3$ zigzag order is E$_{3rd NN}=J_3 \bf{s_3.(-s_3})=$-0.5 meV and the energy at the domain wall is E$_{3rd NN}^{DW}=J_3 \bf{s_3.(-s_1})\simeq$-0.0597 meV. Energy cost due to third nearest neighbour at domain wall is $\simeq-0.0597-(-0.5)\simeq0.4403$ meV. Therefore, the total cost due to first and third nearest neighbours per unit length is $(5.2223+0.4403)/\sqrt{3}a_0=5.6623/\sqrt{3}a_0$ where $a_0$ is the bond length ($a/\sqrt{3}$) of the honeycomb lattice.

\section{Details of the atomistic simulations} \label{app:LLG}
In order to determine the lowest ground state of the Hamiltonian, we solve the Landau-Lifshitz-Gilbert (LLG) equations: \cite{1353448}
\begin{eqnarray} \label{eq:E1}
\frac{d\textbf{S}}{dt}=-\gamma \textbf{S}\times \textbf{B}^{\rm eff}+\alpha \textbf{S} \times \frac{d\textbf{S}}{dt} \, ,
\end{eqnarray}
where $\textbf{B}^{\rm eff}=-\delta H/\delta \textbf{S}$ is the effective magnetic field, $\gamma$ is the gyromagnetic ratio and $\alpha$ is Gilbert damping coefficient. We solve the LLG equations in a self-consistent manner, imposing the constraint of $|{\bf{S}}|=1$ and applied periodic boundary conditions. We use the semi-implicit midpoint method \cite{Mentink_2010} in MATLAB software. We considered multiple random and ferromagnetic spin configurations as the initial state for a particular magnetic field and twist angle and picked the lowest energy configuration after convergence.


%